\documentclass[showpacs,preprintnumbers,amsmath,amssymb,twocolumn,aps,pre]{revtex4}

\usepackage{graphicx}

\begin{document}

\title{Resonance induced by repulsive interactions in a model of globally-coupled bistable systems}
\author{T. Vaz Martins$^{1,2}$, V. N. Livina$^{3}$, A. P. Majtey$^{4}$, R. Toral$^{1}$}

\affiliation{(1)IFISC, Instituto de F{\'\i}sica Interdisciplinar y Sistemas Complejos, CSIC-UIB,  Campus UIB, E-07122 Palma de Mallorca, Spain \\(2)Centro de F{\'\i}sica do Porto, DF, FCUP, 4169-007 Porto, Portugal \\(3)School of Environmental Sciences, University of East Anglia, Norwich, NR4 7TJ, UK \\(4) Departament de F{\' \i}sica, Universitat de les Illes Balears, E-07122 Palma de Mallorca, Spain }

\date{\today}


\begin{abstract}

We show the existence of a competition-induced resonance effect for a generic globally coupled bistable system. In particular, we demonstrate that the response of the macroscopic variable to an external signal is optimal for a particular proportion of repulsive links. Furthermore, we show that a resonance also occurs for other system parameters, like the coupling strength and the number of elements. We relate this resonance to the appearance of a multistable region, and we predict the location of the resonance peaks, by a simple spectral analysis of the Laplacian matrix.

\end{abstract}

\maketitle

\section{Introduction}
\label{intro}

The amplification of an external forcing acting upon a dynamical system under the presence of the right amount of disorder has attracted much attention in the last decades. The phenomenon of {\sl stochastic resonance}, proposed in 1981 to explain the periodicity of ice ages,~\cite{benzi,nicolis} is a somehow counterintuitive effect arising from the cooperation between deterministic dynamics and dynamical disorder or noise. By this effect, a system's coherent response to a weak signal can be optimally amplified by an intermediate level of noise. The prototypical example is that of a continuous variable whose deterministic dynamics is relaxational in a double-well potential. Noise induces jumps between the wells with a rate given by Kramers' expression~\cite{kramers}. The system becomes optimally synchronized with the signal when the signal half-period matches Kramers' rate, as reflected by a maximum value in a suitably defined response. Applications of stochastic resonance wer
 e addressed in many areas, and the theory evolved in several directions (see~\cite{G} and~\cite{HM} for reviews). We would like to stress here the focus on many-component, or extended, systems~\cite{extend, benzic} for which it was shown that it is possible to tune some of the parameters, like the number of elements~\cite{SSSR, tessone, wio, schmidt}, or coupling strength~\cite{JBPM1992,MGC1995,gang,jung,lindner,schi,neiman,us06,us08}, to further enhance the resonance effect. 

Another, more recent, related line of research considers the role that other types of disorder, such as quenched noise (identified with heterogeneity or disorder), can play in producing a resonance effect in many-component systems. Tessone {\sl et al.}~\cite{TMTG06, TTVL} have shown that in generic bistable or excitable systems, an intermediate level of diversity in the individual units can enhance the global response to a weak signal. In bistable systems, a mean field analysis interprets the phenomenon macroscopically as a result of the two equilibrium states getting closer to each other and the lowering of the height of the potential barrier, thereby turning the signal suprathreshold. The resonance appears close to an order-disorder transition, and it was suggested that any source of disorder would lead to the same result. This {\sl diversity-induced resonance} effect has already found extensions in systems concerned with cellular signaling~\cite{chen2007}, complex networks
 ~\cite{acebron2007}, intercellular Ca$^{2+}$ wave propagation induced by cellular diversity~\cite{gosak}, linear oscillators \cite{linosc} and others. Focusing on the double-well model, Perc {\sl et al.}~\cite{Perc} studied the combined effect of dynamic and static disorder, where static disorder was either diversity, the presence of competitive interactions, or a random field. Namely, they showed that the random presence of repulsive bonds decreases the level of noise warranting the optimal response. 

It is the purpose of this work to show that competitive interactions can actually replace -not merely enhance- noise in its constructive effect. In a recent study~\cite{dac} we focused on a network of Ising-like units, and found that the addition of an intermediate fraction of repulsive links can increase the sensitivity to an external forcing. In this work, we expand the result by looking at a generic double-well model, and clarifying some of the previous conclusions. We resort to a spectral analysis of the Laplacian \cite{pat} matrix, to locate the amplification region, and unveil the mechanism of resonance.

The outline of this paper is as follows: in section \ref{themodel} we will introduce the model; we show that there is an amplification and discuss how the amplification mechanism is related to a break of stability in section \ref{response}; and how we can predict the resonance peaks in section \ref{spectral}; Conclusions are drawn in section \ref{conclusions}.

\section{The model}
\label{themodel}

We consider a system of $N$ globally-coupled bistable units described by real variables $s_i(t), i=1,\dots,N$ under the influence of a periodic forcing.
\begin{equation}
\frac{ds_i}{dt}=s_i-s_i^3+\frac{C}{N}\sum_{j=1}^N J_{ij}(s_j-s_i)+A\sin(2\pi t/T),
 \label{phi4}
\end{equation}
where $t$ is the dimensionless time, $C$ measures the coupling strength amongst the different units and $A\sin(2\pi t/T)$ is a periodic external signal with amplitude $A$ and period $T$. The interaction matrix $J_{ij}$ reflects the presence of attractive and repulsive interactions between the units. More specifically, we adopt the following values at random:
\begin{equation}
J_{ij}=J_{ji}= \begin{cases}
 -1, & \text{with probability $p$},\\
 1, & \text{with probability $1-p$}.
\end{cases}
\label{Eq:Jij}
\end{equation}
The single-element case, $N=1$, with added noise is the prototypical double-well potential system for which stochastic resonance was first considered. The case without repulsive interactions, $p=0$, can still be described globally by a bistable potential (see next section) and, in the presence of noise, has been widely studied as a model case for stochastic resonance in extended systems~\cite{benzic}; it has also been considered,  in the presence of a random field, as a prototypical example for the diversity-induced resonance effect~\cite{TMTG06}.  For $p>0$, the coexistence of attractive and repulsive interactions is characteristic of a wide class of spin-glass-type systems~\cite{spinglass2}.

We will focus on the macroscopic variable $S(t)=\frac{1}{N}\sum_i s_i(t)$, and use as a measure of response the spectral power amplification factor~\cite{ampli}, defined as the ratio of the output to input power at the corresponding driving frequency: 
\begin{equation}
R=4A^{-2}\left|\langle{\rm e}^{-i2\pi t/T}S(t)\rangle\right|^2
\label{eq:R}
\end{equation}
where $\langle \cdots\rangle$ is a time average. $R$ is roughly proportional  to the amplitude of the oscillations of $S(t)$. If $R<1$, then the amplitude of the response is less than that of the signal, and vice versa for $R>1$. 

\section{Signal amplification}
\label{response}

It is convenient to analyze first the structure of the steady-state solutions for the system of equations (\ref{phi4}) in the non-forced case, $A=0$. The dynamics is relaxational $\displaystyle\frac{ds_i}{dt}=-\frac{\partial V}{\partial s_i}$~\cite{ST:1999}, being 
\begin{equation}
V(s_1,\dots,s_N)=\sum_{i=1}^N\left[-\frac{s_i^2}{2}+\frac{s_i^4}{4}+\frac{C}{4N}\sum_{j=1}^N J_{ij}(s_i-s_j)^2\right]
\end{equation}
the Lyapunov potential. Therefore, the stable steady states are the configurations $(s_1,\dots,s_N)$ which are absolute minima of $V$. If there are no repulsive links, $p=0$, the Lyapunov potential has just two equivalent minima at $s_i=+1$ or $s_i=-1$, $\forall i=1,\dots,N$ and, hence, the macroscopic variable will reach the stable asymptotic values $S=+1$ or $S=-1$, a typical situation of bistability. Whether $S=+1$ or $S=-1$ is reached depends exclusively on the initial conditions. As $p$ increases, the absolute minima depart from $S=\pm 1$ and, furthermore, new metastable minima of $V$ appear. The dynamical equations (\ref{phi4}) may or may not get stuck in one of these minima, depending on initial conditions and the particular realization of the coupling constants $J_{ij}$. We have used throughout the paper random initial conditions drawn from a uniform distribution in the $(-1,1)$ interval, although we have observed the same type of phenomenology when using other random
 , but still symmetric, distributions such as truncated Gaussian or the Johnson family of distributions.

\begin{figure}
\begin{center}
\includegraphics[scale=0.75,angle=0,clip]{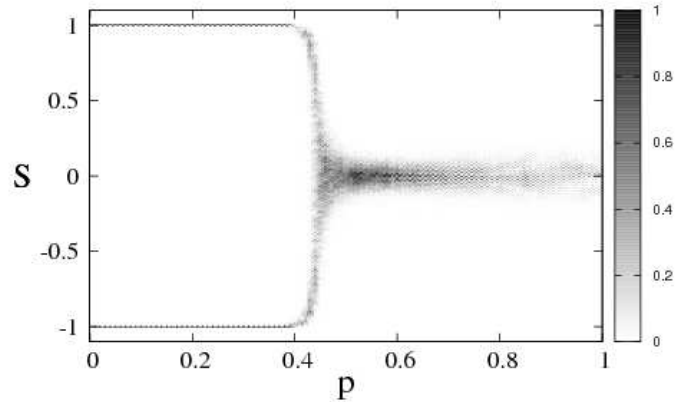}
\caption{We plot in a gray scale the stationary probability distribution $P(S)$, in the absence of external signal $A=0$, coming from numerical simulations of Eqs.~(\ref{phi4}).  For better viewing, the distribution has been rescaled by its maximum value at each $p$.  The data show that at $p<p_c\approx 0.44$ the system presents two equivalent absolute maxima for $P(s)$, while there is only one absolute maximum for $p>p_c$. We note, however, that there are many relative maxima for all values of $p$, specially around the region $p\approx p_c$.  Other parameter values are: $N=200$, $C=8$. The probability has been computed after averaging over $1000$ realizations of the couplings $J_{ij}$ and initial conditions drawn from an uniform distribution in the interval $(-1,1)$. For the numerical integration we used a fourth-order Runge-Kutta method with a time step $\Delta t=0.1$.
\label{ss}}
\end{center}
\end{figure}

>From our simulations we compute numerically the probability distribution $P(S)$ of the final values of $S$ reached during the dynamical evolution for different realizations of the coupling constants $J_{ij}$ and initial conditions. This is plotted in Fig.~\ref{ss}. We can observe a second-order phase transition as the average value $\langle |S(t)|\rangle$ vanishes for $p>p_c\approx 0.44$. One can interpret these results in terms of an effective potential $V_{\text{eff}}(S)\equiv -\ln P(S)$ which has two equivalent absolute minima at $S= \pm S_0(p)$, where $1>S_0(p)>0$ for $0<p<p_c$, and one absolute minimum at $S=0$ for $p\ge p_c$. The effective potential $V_{\text{eff}}$ presents many relative minima for all values of $p>0$, especially in the critical region $p\approx p_c$, a typical situation for the spin-glass models~\cite{spinglass2}.

\begin{figure}
\begin{center}
\includegraphics[scale=0.35,angle=0,clip]{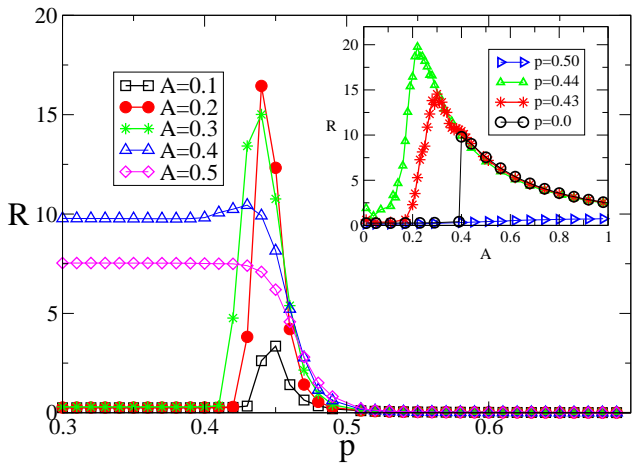}
\caption{Spectral amplification factor $R$ versus probability of repulsive links $p$. Inset: the influence of the amplitude $A$ of the external forcing on the response $R$.  For suprathreshold amplitudes, $A\gtrsim 0.4$,  $R$ decreases with $A$ due to the denominator $A^2$ in the definition of the spectral amplification factor $R$. $T=300$, $N$ and $C$ as in Fig.~\ref{ss}.
\label{a}}
\end{center}
\end{figure}

We now turn on the forcing $A>0$ and study the system response, as measured by the spectral power amplification factor $R$ defined above, Eq.~(\ref{eq:R}). Consider first the case $p=0$. For a small, sub-threshold, amplitude $A$ the macroscopic variable $S(t)$ will just execute small oscillations of amplitude proportional to $A$ around the stable values $S=+1$ or $S=-1$. As $A$ increases beyond the threshold value $A_o\approx 0.4$ the amplitude of the forcing is large enough to induce large jumps of the macroscopic variable from $S\approx -1$ to $S\approx +1$ and vice versa. This change of behavior at $A_o$ appears as a sudden increase in the value of $R$, as shown in the inset of Fig.~\ref{a}. As the same inset shows, similar behavior is observed for $0<p\lesssim p_c$: the response shows a sudden increase for a particular value of the amplitude $A$ and then decreases monotonically. For $p>p_c$, the response is very small and almost independent on the value of $A$.

\begin{figure}
\begin{center}
\includegraphics[scale=0.35,angle=0,clip]{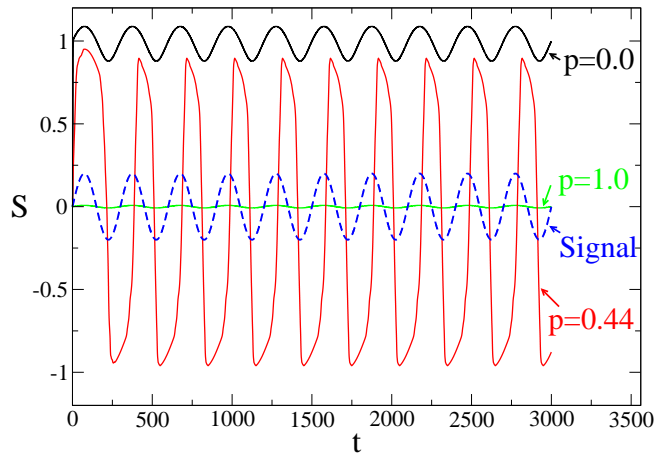}
\caption{Representative trajectories of the macroscopic variable $S(t)$.  Note the large amplitude of the oscillations in the intermediate case $p=0.44$. The ``signal'' is the periodic function $A\sin(2\pi t/T)$. Values of $N$, $C$ and $T$ as in Fig.~\ref{a}, $A=0.2$.
\label{tra}}
\end{center}
\end{figure}

More interesting, and the main result of this paper, is the dependence of $R$ on the probability $p$ of repulsive links, main plot in Fig.~\ref{a}. We note that there is an optimal probability of repulsive links that is able to amplify signals whose amplitude would be sub-threshold in the case $p=0$, i.e. $A<A_o$. For suprathreshold signals, $A>A_o$, the presence of repulsive links does no longer lead to enhanced amplification. As shown in the figure, the optimal value for amplification is close to the critical value $p_c$ signaling the transition from bistability to monostability in the non-forced case. The optimal amplification as a function of $p$ can clearly be observed in Fig.(\ref{tra}) which shows representative trajectories for $p=0$ (small oscillations around the value $S=+1$), $p=p_c$ (large oscillations between $S\approx +1$ and $S\approx -1$) and $p=1$ (small oscillations around $S=0$).

\begin{figure}
\begin{center}
\includegraphics[scale=0.35,angle=0,clip]{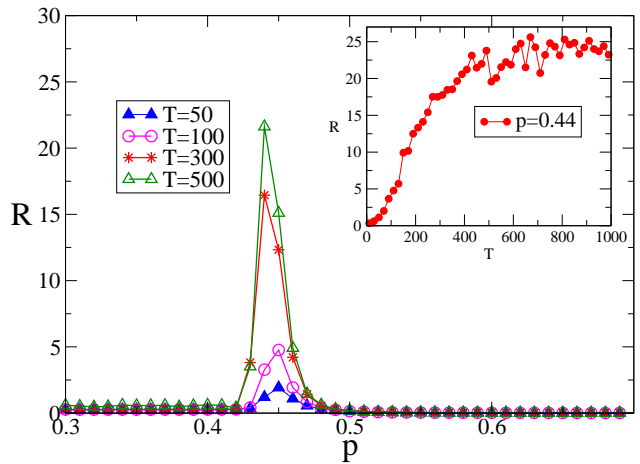}
\caption{The influence of the signal period $T$ on the response $R$. In the inset, we see the response $R$ reaches a constant value for slow enough signals.  Values of $N$, $C$ and $A$ as in Fig.~\ref{tra}.
\label{t}}
\end{center}
\end{figure}

The existence of an optimal value of the fraction of repulsive $p$ for which signal amplification is maximum is somehow reminiscent of the stochastic resonance phenomenon. There are some important differences, however. While in stochastic resonance, the response $R$ shows a maximum as a function of period, resulting from the matching between Kramers' rate and the forcing half-period, in our case the same optimal disorder $p$ amplifies responses to signals of every period, as shown in Fig.~\ref{t}. When the signal is slow enough, the system has time to respond to the fuller extent, going to the absolute extrema of the potential, and the amplification factor reaches a constant value, see inset of Fig.~\ref{t}. 

\begin{figure}
\begin{center}
\includegraphics[scale=0.35,angle=0,clip]{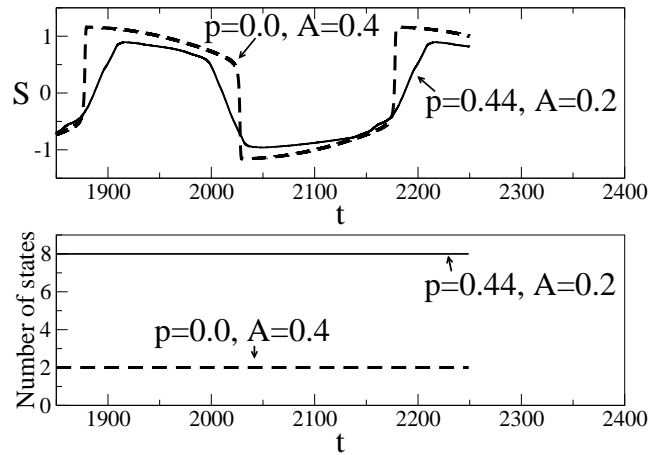}
\caption{We amplify some representative trajectories (upper panel), and count the number of states through which the system moves in each trajectory (lower panel).  Values of $T$, $N$ and $C$ as in Fig.~\ref{a}. \label{ns}}
\end{center}
\end{figure}

It is possible to reinterpret these results in terms of the effective potential $V_{\text{eff}}(S)$ introduced above. The periodic forcing can be seen, approximately, as a periodic modulation to the potential $V_{\text{eff}}-S\cdot A\sin(2\pi t/T)$. As discussed above, the effect of the repulsive links is such that $V_{\text{eff}}(S)$ changes from bistable at $p=0$ to having many metastable minima at $p\approx p_c$ and a single absolute minimum for $p>p_c$. Hence, the deep potential barrier separating the $S=\pm 1$ solutions for $p=0$ lowers under the effect of the repulsive links. As a consequence, the modulation induced by the periodic forcing is now large enough, and the global variable is then able to oscillate from the minimum $+S_0(p)$ to $-S_0(p)$ and vice versa. As $p$ approaches $p_c$ a more complicated scenario appears. In this region the effective potential presents already a rich structure with many metastable minima in the non-forced case. Those minima can be mod
 ified or even disappear by the effect of the periodic modulation. It is particularly illustrative to compare the responses to a suprathreshold signal of amplitude $A=0.4$ in the case $p=0$,  and to a signal of amplitude $A=0.2$ (which would be subthreshold in the case $p=0$) at the optimal fraction of repulsive links $p=p_c$. In both cases, the amplitude of the oscillations is approximately the same, as the system makes large excursions from $S\approx -1$ to $S\approx +1$ and vice versa. However, the shape of the oscillations is rather different, as shown in the upper panel of Fig \ref{ns}. In the $p=0$ case, the transition from one minimum to the other is rather fast (vertical portion of the dashed line), while in the case $p=p_c$, the transition is slower as the system seems to be spending more time in intermediate states.

To determine what those differences reveal about the underlying effective potential, we have used a method~\cite{Liv} that allows us to detect the number of states a system visits from an analysis of its time series. A typical example is shown in the lower panel of Fig.~\ref{ns}. We only detect two states in the global variable $S$ when $p=0$, corresponding, as expected, to the modulated bistable potential. By contrast, the slight irregularities in the trajectory for $p=0.44$, hardly visible by eye, correspond to several very shallow potential wells. The system evolves through many states at the optimal probability of repulsive links, as shown in the lower panel of Fig~\ref{ns}. This image explains why signals of every amplitude and period can be amplified for $p\approx p_c$. In this case, the system can access the many intermediate states, covering a distance proportional to $T$ and $A$, in case of very fast or very weak signals. 

The previous results show that the disorder induced by an intermediate level of repulsive links is an essential ingredient to get an optimal response to the external forcing. This can be explained as, in the absence of forcing, the metastable states correspond to a wide distribution for the values of $s_i$'s. When the forcing is turned on, some units will be responsive to the signal, and then they will pull others which are positively coupled to them. This basic mechanism is further highlighted by the observation of a resonance behavior with both the coupling constant $C$ and the number of units $N$.  

\begin{figure}
\begin{center}
\includegraphics[scale=0.35,angle=0,clip]{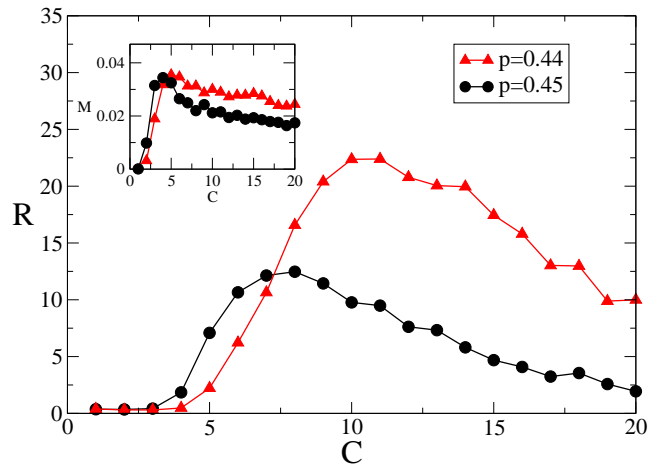}
\caption{Coupling-induced resonance. Main plot: the response $R$ shows a maximum as a function of coupling constant $C$.
As shown in the insert, the same maximum appears as a function of the localization measure $M$, see section \ref{spectral}.   Values of $N$, $T$ and $A$ as in figure \ref{tra} and $K=0.2$ in the insert.
\label{cm}}
\end{center}
\end{figure}

\begin{figure}
\begin{center}
\includegraphics[scale=0.35,angle=0,clip]{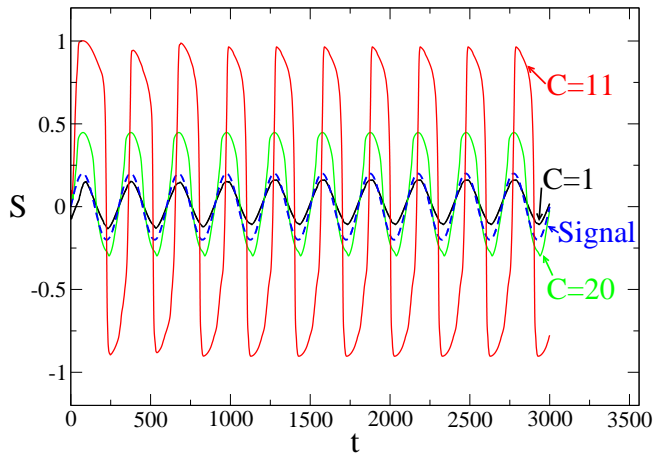}
\caption{Coupling-induced resonance, as revealed by the resonant trajectory at optimal $C=11$. Values of $N$, $A$ and $T$ as in Fig.~\ref{tra}.
\label{sc}}
\end{center}
\end{figure}

The resonance with $C$ and some representative trajectories are displayed in Figs.~\ref{cm} and \ref{sc}, respectively. In the weak coupling limit, the units behave basically as independent from each other and, as the signal amplitude $A$ is subthreshold for a single variable, the overall response is small. In the large coupling limit, the interaction term is too big to allow an unit that could first follow the signal to depart from the influence of its neighbors.

\begin{figure}
\begin{center}
\includegraphics[scale=0.35,angle=0,clip]{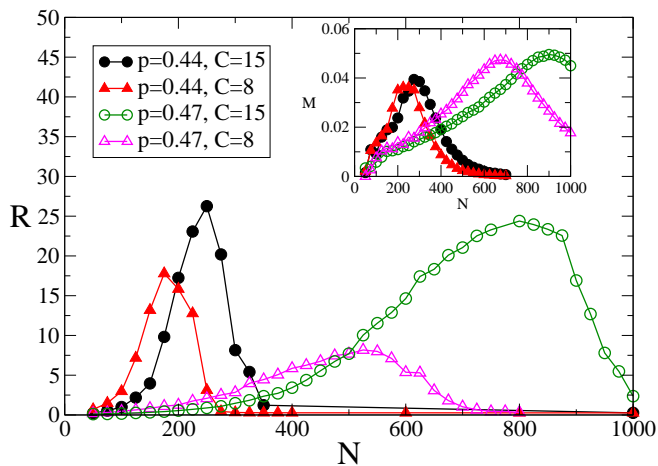}
\caption{System-size induced resonance. Main plot: the response $R$ shows a maximum as a function of the number of units, that follows the same pattern as the maximum $M$ ($K=0.2$) (inset). Since $N$ decreases the influence of a single neighbor, and $C$ increases it, when the coupling intensity is larger, the optimal system size increases. Values of $T$ and $A$ as in Fig.~\ref{tra}.
\label{nm}}
\end{center}
\end{figure}

\begin{figure}
\begin{center}
\includegraphics[scale=0.35,angle=0,clip]{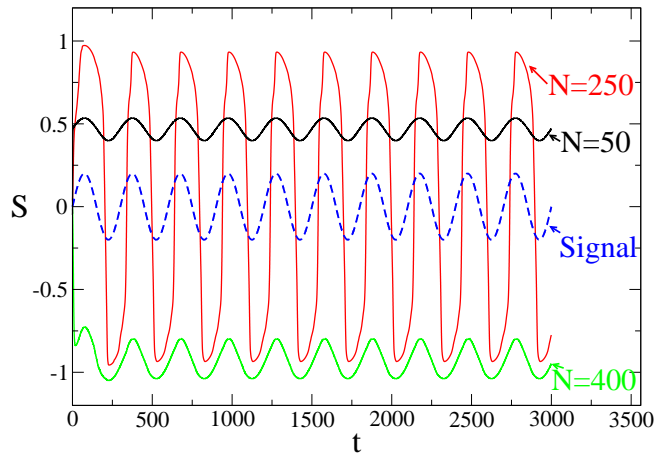}
\caption{System size induced resonance, as revealed by the resonant trajectory at optimal $N=250$. Values of $T$, $A$ and $C$ as in Fig.~\ref{tra}.
\label{tn}}
\end{center}
\end{figure}

The resonance with the number of units $N$ and some representative trajectories is presented in Figs.~\ref{nm} and \ref{tn}. Since fluctuations in the number of repulsive links decrease with $N$, a larger system requires a greater fraction of repulsive links to achieve the same level of disorder than a smaller system. As a consequence, the response of a larger system is best amplified at a higher probability of repulsive links. As the fraction of repulsive links must not exceed the fraction of positive ones, there can be a limit on how large can a system be, to be able to amplify a signal. The same behavior, focusing on the number of neighbors was found in a previous study of an Ising-like network model~\cite{dac}.
 
\section{Spectral analysis}
\label{spectral}

We have already commented that the optimal probability of repulsive links drives the system to a glassy phase. Anderson \cite{and58, and70} has proposed a connection between a glass and a delocalization-localization transition, relating the existence of many metastable states with a localization of modes. From this proposal, we retain the idea to work in the eigenspace of the interaction matrix, and to look for the fraction of repulsive links where mode localization becomes significant. This approach has the virtue of not only identifying the steady states, but also to shed light onto how the reaction to perturbations is sustained and spreads along the system, depending on the fraction of repulsive links. In this manner, we hope to locate the region where multistability is expected, and also to understand the mechanism of response to external perturbations.

\begin{figure}
\begin{center}
\includegraphics[scale=0.40,angle=0,clip]{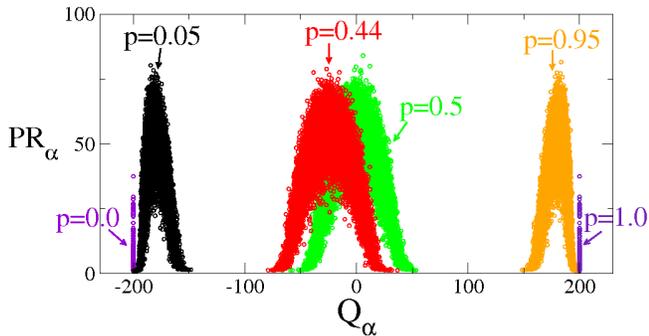}
\caption{The participation ratio $\text{PR}_\alpha$ is a measure of localization; it estimates the number of eigenvectors components contributing to the $Q_\alpha$ eigenvalue. The eigenvalues at both ends of the spectrum are localized for intermediate levels of disorder. $N=200$,  results after 1000 independent runs. \label{spe}}
\end{center}
\end{figure}

Following \cite{Perc}, let us define the eigenvalues $Q_{\alpha}$ and (normalized) eigenvectors $e^{\alpha}=(e_1^{\alpha},\dots,e_N^{\alpha})$ of the Laplacian matrix \cite{pat} $J'_{ij}$ :
\begin{equation}
J'_{ij}=J_{ij}-\delta_{ij}\sum_{k=1}^N J_{kj},
\end{equation}
\begin{equation}
\sum_{j=1}^NJ'_{ij}e_j^{\alpha}=Q_{\alpha}e_i^{\alpha}.
\end{equation}
The effect of the competitive interactions can be described by the so-called participation ratio of eigenvector $e^{\alpha}$, defined as $\text{PR}_\alpha=1/\sum_{i=1}^N [e_i^{\alpha}]^4$. It quantifies the number of components that participate significantly in each eigenvector. A state $\alpha$ with equal components has $\text{PR}_\alpha=N$, and one with only one component has $\text{PR}_\alpha=1$. When $\text{PR}_\alpha=1$ on a fraction $f$ of elements, and $0$ elsewhere, then $\text{PR}_\alpha=f$, which justifies its name.
More precisely, we will define ``localized'' modes as the ones whose participation ratio is less than $0.1N$. Our first observation (Fig.~\ref{spe}) is that at the optimal region $p$ there is a significant fraction of positive eigenvalues, and, of those, a significant fraction of the corresponding eigenstates are localized. In this region, we will neglect the coupling between different modes. This approximation allows us to look in more detail at what happens at the optimal region, and in particular at the effect of the coupling strength $C$ and the number $N$ of elements. 
 
Let us focus first on the unforced system ($A=0$), to see how the presence of the disorder induced by the repulsive links affects a state configuration. We assume each unit $i$ is initially at a given state $s_i^o$, chosen from a random symmetric distribution and split the variables in the steady state as $s_i=s_i^o+x_i$, being $x_i$ the deviation from the initial condition. We express $x_i$ in the eigenbasis of the $J'_{ij}$ matrix:
\begin{equation}
x_i=\sum_{\alpha=1}^NB_{\alpha}e_i^{\alpha},
\end{equation}

Expanding Eq. (\ref{phi4}) for $A=0$, multiplying the resulting equation by $e_i^{\alpha}$, summing over all elements $i$, and approximating averages of the product of initial conditions and eigenvectors by the product of their individual averages (e.g. $\sum_{i=1}^Ns_i^oe_i^{\alpha}\approx \sum_{i=1}^Ns_i^o\sum_{i=1}^Ne_i^{\alpha}=0$), we obtain:
\begin{equation}
\sum_{\beta,\gamma,\eta}
F^{\beta\gamma\eta\alpha}B_{\beta}B_{\gamma}B_{\eta}
+\left(K-C\frac{Q_{\alpha}}{N}\right)B_{\alpha}=0,
\end{equation}
where
\begin{eqnarray}
F^{\beta\gamma\eta\alpha}&=&\sum_{i=1}^N
e_i^{\beta}e_i^{\gamma}e_i^{\eta}e_i^{\alpha},\\
 \label{k}
K&= &\frac{3}{N} \sum_{i=1}^N (s_i^o)^2 -1.
\end{eqnarray}
Neglecting coupling between modes leads to $F^{\beta\gamma\eta\alpha}=1/\text{PR}_{\alpha}$ if $\alpha=\beta=\gamma=\eta$ and $F^{\beta\gamma\eta\alpha}=0$ otherwise. We then obtain the following equation for the amplitude of the $\alpha$-th mode:
\begin{equation}
B_{\alpha}^3 +\text{PR}_{\alpha}\left(K-C\frac{Q_{\alpha}}{N}\right)B_{\alpha}=0.
\end{equation}
According to this approximation, unless $Q_{\alpha}>\frac{K N}{C}$, the amplitude $B_{\alpha}$  of the mode $\alpha$ is zero, and any small perturbation vanishes.  Otherwise, the mode is said ``open'' and $B_{\alpha}$ takes one of the values:
\begin{equation}
\label{aalfa}
B_{\alpha}=\pm \sqrt{\text{PR}_{\alpha}\left(C\frac{Q_{\alpha}}{N} -K\right)}.
\end{equation}
For intermediate amounts of disorder, some open modes begin to appear. The final state of an unit is $s_i = s_i^o + \sum_{\alpha=1}^NB_{\alpha}e_i^{\alpha}$, and when the initial conditions  are random and the open modes $\alpha$ are localized, the system reaches many metastable states, given all the possible combinations of individual states. For this reason, we want to locate a transition to a region with a significant number of localized modes. 

To concretize, we define a measure $M$ of localization: 
\begin{equation}
M=\frac{N_L^2}{N_O N},
\end{equation}
where $N_O$ is the number of modes $\alpha$ whose associated eigenvalue $Q_{\alpha}$ is greater than $\frac{K N}{C}$, and $N_L$ is the number of those modes which, in addition, are localized, i.e. $\text{PR}_{\alpha}<0.1N$. 

\begin{figure}
\begin{center}
\includegraphics[scale=0.35,angle=0,clip]{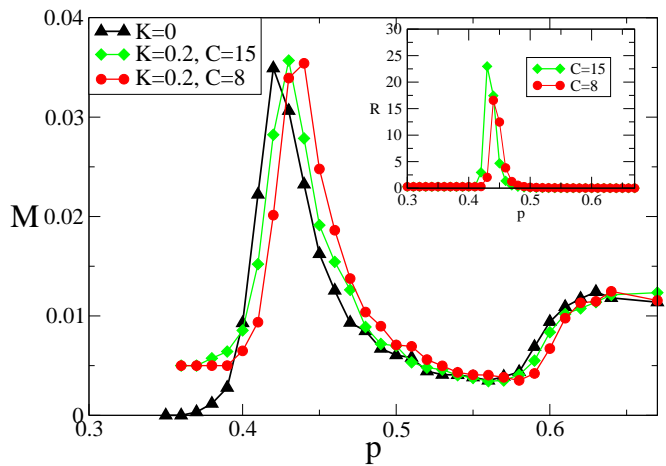}
\caption{ Measure of localization M: close to the optimal region there is an increase in the fraction of localized open eigenstates. The small dependance of $M$ with $K$ and $C$ is as expected.  Parameters: $N=200$. Inset: $A=0.2$, $T=300$.\label{pm}}
\end{center}
\end{figure}

Recalling the definition of $K$ (Eq.~(\ref{k})), we see its value is related to a choice of initial conditions, by the variance of $s_i^o$.   Since we expect multistability to emerge when the initial distribution is more or less uniform, we present results in Fig.~\ref{pm} for values of $K\approx 0$.

 At moderate levels of disorder, the localized nodes appear on the tails of the spectra, Fig. \ref{spe}. We confirm that the optimal probability of repulsive links coincides with a maximal localization of open modes in that region, as identified by the peak in $M$ (Fig.~\ref{pm}). 

In a particular metastable state, the units are randomly distributed, more concentrated near one of the potential wells. Observing the results in Fig.~\ref{pm} for $K\gtrsim 0$, we see that we recover the dependance on $C$, and that the peak in $M$ still coincides with the optimal probability region. The enhanced responsiveness to an external signal can thus be understood as a consequence of mode localization. Since units can be  in  different positions, some will be able to answer the signal, and then -  since the overall coupling is attractive - pull the others. This is done in an incremental fashion, as confirmed by the localized reaction to perturbations. 

The same analysis is valid when we plot $M$ as a function of $C$ or $N$. We notice a peak in $M$ and accordingly the dependance of the response on $C$ (Fig.~\ref{cm}) and $N$ (Fig.~\ref{nm}) shows a maximum for intermediate values (insets Fig.~\ref{cm} and Fig.~\ref{nm}).
When $C$ is small, even if the modes are open, their amplitude $B_{\alpha}$ is weak, Eq.~(\ref{aalfa}). A high fraction of repulsive links, increasing the number of open modes, can overcome this situation to a certain degree, allowing for resonances at a smaller coupling strength. 
 
\section{Conclusions}
\label{conclusions}

In this work, we have analyzed the response to a weak period signal, of a model composed by bistable units coupled through both attractive and repulsive links. 

Our main result is that the system collective response is enhanced by the presence of an intermediate fraction of repulsive links. Hence, competitive interactions are taken as a source of disorder, as an alternative to previous studies where disorder was induced by noise~\cite{G} or diversity~\cite{TMTG06}, and a similar amplification was verified.
 
We have chosen a very generic double-well model, and have shown that the optimal disorder is the one that destroys the ordered system bistability. The resulting multistable effective potential allows for the amplification of very weak or fast signals. There is not a need to match specific levels of disorder with specific frequencies. Having the optimal disorder, the system becomes more sensitive to external signals of every kind. Furthermore, we have shown that varying the number of elements or the coupling strength in an ensemble of coupled bistable elements can improve the sensitivity to an external forcing. These various ways to increase sensitivity make the phenomenon less dependent on a fine tuning of the proportion of repulsive links, which can be a positive feature in practical applications. Apparently, when the system size becomes very large, it is difficult to get a resonance effect, unless we increase the coupling strength by many times. Arguably, thi
 s difficulty can be overcome by other types of network settings~\cite{TZT08}.

Finally, we have shown that the location of the resonance peaks can be predicted by a spectral analysis of the Laplacian matrix. In heuristic terms \cite{pat}, the positive eigenvalues of the Laplacian can be seen to express the contribution of the coupling term to the vulnerability of the system to perturbations. We conclude that the location of the amplification region, for a given system size and coupling constant, is reasonably independent of the particular dynamical system. In broad terms, it corresponds to the point where the positive eigenvalues of the Laplacian matrix become localized, signaling a transition to a region where perturbations can accumulate in an incremental manner. The more precise location would depend on the particular dynamical system by means of a condition on open modes.

Competitive interactions are widespread in nature, notably in biological systems. In those systems and others, there has been some studies highlighting their role in achieving a coherent behavior in the absence of forcing: increasing synchronization \cite{Levya} or enabling a collective firing~\cite{TZT08}. In the present study, we saw they can also help to enhance perception, something that can be potentially relevant in sensory systems.

\bigskip
{\textbf{Acknowledgments:}}
We acknowledge financial support by the MEC (Spain) and FEDER (EU) through project FIS2007-60327. TVM is supported by FCT (Portugal) through Grant No.SFRH/BD/23709/2005. VNL acknowledges financial support of NERC (project NE/F005474/1) and postdoctoral fellowship of the AXA Research Fund.

\end{document}